\documentclass[traditabstract,longauth]{aa}  
\usepackage{graphicx}
\usepackage[rightcaption]{sidecap}

\usepackage{txfonts}

\usepackage{multirow}

\usepackage{svg}
\usepackage{hyperref}
\usepackage{placeins}
\hypersetup{colorlinks,citecolor=blue}
\newcommand{\pycasso}{{\sc p}y{\sc casso}}          
\newcommand{\starlight}{{\sc starlight}}          	
\newcommand{\shifu}{{\sc shifu}}                 	
\usepackage{balance}
\usepackage{float} 
\usepackage{stfloats}

\usepackage{longtable}
\usepackage{afterpage}
 
\newcommand{\nii}{\ion{N}{II}}
\newcommand{\oiii}{\ion{O}{III}}

\begin{document} 

\titlerunning{The CAVITY project. The spatially resolved SFR of galaxies in voids}
\title{The CAVITY project. The spatially resolved SFR of galaxies in voids}

\author{Ana M. Conrado \inst{\ref{inst:IAA}}
\and Rub\'en~Garc\'ia-Benito\inst{\ref{inst:IAA}}
\and Rosa~M.~Gonz\'alez Delgado\inst{\ref{inst:IAA}}
\and Bahar~Bidaran\inst{\ref{inst:UGR}}
\and H\'el\`ene M.~Courtois\inst{\ref{inst:Lyon}}
\and Salvador~Duarte Puertas\inst{\ref{inst:UGR}, \ref{inst:CarlosI}, \ref{inst:Quebec}}
\and Daniel~Espada\inst{\ref{inst:UGR}, \ref{inst:CarlosI}}
\and Andoni~Jim\'enez\inst{\ref{inst:UGR}}
\and Ignacio~del Moral-Castro\inst{\ref{inst:Chile}}
\and Isabel~P\'erez\inst{\ref{inst:UGR}, \ref{inst:CarlosI}}
\and Tom\'as~Ruiz-Lara\inst{\ref{inst:UGR}, \ref{inst:CarlosI}}
\and Laura~S\'anchez-Menguiano\inst{\ref{inst:UGR}, \ref{inst:CarlosI}}
\and Gloria~Torres-R\'ios\inst{\ref{inst:UGR}}
\and Simon~Verley\inst{\ref{inst:UGR}, \ref{inst:CarlosI}}
\and Mar\'ia~Argudo-Fern\'andez\inst{\ref{inst:UGR}, \ref{inst:CarlosI}}
\and Simon B.~De Daniloff\inst{\ref{inst:IRAM},\ref{inst:UGR}}
\and Estrella~Florido\inst{\ref{inst:UGR}, \ref{inst:CarlosI}}
\and Yllari K.~González-Koda\inst{\ref{inst:UGR}}
\and Alejandra Z.~Lugo-Aranda\inst{\ref{inst:Mexico}}
\and Javier~Rom\'an\inst{\ref{inst:Cordoba}}
\and Smitha~Subramanian\inst{\ref{inst:India}, \ref{inst:IAP}}
\and Pedro~Villalba-Gonz\'alez\inst{\ref{inst:Vancouver}}
\and Manuel~Alc\'azar-Laynez\inst{\ref{inst:UGR}}
\and M\'onica~Hern\'andez-S\'anchez\inst{\ref{inst:Valencia}}
\and M\'onica~Rodr\'iguez Mart\'inez\inst{\ref{inst:IRAM}}
\and Paulo~V\'asquez-Bustos\inst{\ref{inst:UGR}}
\and Martin~Blazek\inst{\ref{inst:CAHA}}
}

\institute{Instituto de Astrof\'isica de Andaluc\'ia (CSIC), PO Box 3004, 18008 Granada, Spain \label{inst:IAA} 
\and Departamento de F\'isica Te\'orica y del Cosmos, Facultad de Ciencias (Edificio Mecenas), Universidad de Granada, E-18071 Granada, Spain \label{inst:UGR}
\and Universit\'e Claude Bernard Lyon 1, IUF, IP2I Lyon, 69622 Villeurbanne, France \label{inst:Lyon}
\and Instituto Carlos I de F\'isica Te\'orica y Computacional, Universidad de Granada, E-18071 Granada, Spain \label{inst:CarlosI}
\and D\'epartement de Physique, de G\'enie Physique et d'Optique, Universit\'e Laval, and Centre de Recherche en Astrophysique du Qu\'ebec (CRAQ), Qu\'ebec, QC, G1V 0A6, Canada \label{inst:Quebec}
\and Instituto de Astrofísica, Facultad de Física, Pontificia Universidad Católica de Chile, Campus San Joaquín, Av. Vicuña Mackenna 4860, Macul, 7820436, Santiago, Chile \label{inst:Chile}
\and Institut de Radioastonomie Millim\'etrique (IRAM), Av. Divina Pastora 7, N\'ucleo Central 18012, Granada, Spain \label{inst:IRAM}
\and Instituto de Astronom\'ia, Universidad Nacional Aut\'onoma de M\'exico, A.P. 106, Ensenada 22800, BC, M\'exico \label{inst:Mexico}
\and Departamento de F\'isica, Universidad de C\'ordoba, Campus Universitario de Rabanales, Ctra. N-IV Km. 396, E-14071 C\'ordoba, Spain\label{inst:Cordoba}
\and Indian Institute of Astrophysics, Koramangala II Block, Bangalore-560034, India \label{inst:India}
\and Leibniz-Institut f\"ur Astrophysik Potsdam, An der Sternwarte 16, D-14482 Potsdam, Germany  \label{inst:IAP}
\and University of British Columbia, Vancouver, BC V6T 1Z1, Canada  \label{inst:Vancouver}
\and Departament d'Astronomia i Astrof\'isica, Universitat de Val\`encia, 46100 Burjassot (Val\`encia), Spain \label{inst:Valencia}
\and Centro Astronómico Hispano en Andalucía, Observatorio de Calar Alto, Sierra de los Filabres, 04550 Gérgal, Almería, Spain \label{inst:CAHA}
}

\date{} 

\abstract{
The mass in the Universe is distributed non-uniformly, originating the Large Scale Structure (LSS), characterised by clusters, filaments, walls and voids.
Galaxies in voids have been found to be bluer, of later type, less massive, and to have slower evolution than galaxies in denser large-scale environments. The effect of the void environment on many other evolutionary properties, such as the star formation rate (SFR), is still a topic under discussion. We tackle this problem with a new perspective by estimating the spatially-resolved SFR derived from extinction-corrected H$\alpha$ luminosities of a sample of 220 void galaxies from the  Calar Alto Void Integral-field Treasury surveY (CAVITY) project. These observations consist of optical integral field unit datacubes obtained with the PMAS/PPaK spectrograph at the 3.5 m telescope at the Calar Alto Observatory (Almería, Spain). We measure their continuum-subtracted emission lines to obtain the maps of the SFR, specific star formation rate (sSFR) and extinction ($A_V$). From them, we assess the behaviour of the whole galaxies through their global properties, and the spatially-resolved information with radial profiles up to 2 half-light radii. We then compare the results with a sample of galaxies in filaments and walls from the CALIFA survey analysed using the same methodology. We build a control sample matched in morphological type and total stellar mass by taking pairs of each CAVITY galaxy. We find no statistically significant differences in the SFR and sSFR (p $\leq0.05$), although void galaxies tend to have larger SFR, especially for the early spirals.This effect is present for Sa galaxies at all galactocentric distances ($\Delta$ log sSFR [Gyr$^{-1}$] = 0.25 dex at 1 half light radius, HLR), and can also be seen in the outer parts (R > 1 HLR) of late-type spirals ($\Delta$ log sSFR [Gyr$^{-1}$] = 0.26 dex at 2 HLR), evidencing a slower transition from star-forming to quiescent and less evolved spiral discs. Additionally, we find void late-type galaxies to have lower extinction ($\Delta=0.16\pm0.06$ mag). Using the extinction normalised by the stellar mass surface density as a proxy for the gas mass fraction, we find it to be larger for the void early spirals (by a 5\%), especially in their outer regions (23\% larger). This indicates the effect of the void environment on the transition stage from star forming to passive.}
 
   \maketitle

\section{Introduction} \label{Section: introduction}

The Universe has evolved into the shape of a cosmic web, forming structures such as crowded clusters, worm-like filaments, foliaceous walls, and bleak voids. The location within this structure in which a galaxy forms and evolves has a strong impact on its physical properties. Galaxies that assemble in clusters and groups tend to be redder, elliptical and are forming fewer stars than galaxies in the field \citep{Dressler1980}. On the other end of the density spectrum, galaxies that reside in voids have been found to be fainter and less massive \citep{Moorman2015}, bluer and have later-type morphologies \citep{Rojas2004,Kreckel2011,Hoyle2012,Argudo-Fernandez2024}, and to have slower star formation histories \citep{Dominguez-Gomez2023b} than galaxies located in environments of a higher density.

Differences in the evolutionary properties of void galaxies, such as the star formation rate (SFR), are still under discussion, and many approaches have been followed in the literature. Some studies find that there is an enhancement of the SFR in void galaxies \citep{Rojas2005, Beygu2016A, Moorman2016, Florez2021, Rodriguez-Medrano2023}. Others do not find significant differences between galaxies in voids and comparable galaxies in denser environments \citep{Patiri2006, Ricciardelli2014, Dominguez-gomez2022, Rosas-Guevara2022, Rodriguez2024}. Although these studies use different void samples and SFR tracers, they all analyse galaxies only through their global properties, without exploring their internal spatial variations.

Environmental processes, interactions between galaxies or mechanisms such as gas pressure stripping, can affect star formation in a galaxy by heating or removing gas in its disc \citep{Gunn1972}. These effects have been shown to be stronger in the outer parts of galaxies \citep{Koopmann2004}. Additionally, secular processes have also been found to depend on the galaxies' spatial dimension: massive galaxies have been shown to form and evolve inside-out \citep{Perez2013, Gonzalez-Delgado2014, Gonzalez-Delgado2015, Gonzalez-Delgado2017, Garcia-Benito2017}, and star migration can happen from inner processes, such as the formation of bars. These considerations show that understanding galaxy evolution requires analysing how properties vary spatially within galaxies.

This can be achieved using integral field spectroscopy (IFS), which simultaneously provides information from the spatial and spectral dimensions through 3D data cubes. This way, we can analyse the spectra of the galaxy in different regions, and construct maps of their physical properties. Examples of IFS surveys are ATLAS3D \citep[A Three-dimensional Legacy Survey of Massive Galaxies,][]{Cappellari2011}, CALIFA \citep[Calar Alto Legacy Integral Field Area,][]{Sanchez2012}, SAMI \citep[Sydney-AAO Multi-object Integral-Field Spectrograph,][]{Croom2012}, or MaNGA \citep[Mapping Nearby Galaxies at APO,][]{Bundy2015}.

CAVITY \citep[Calar Alto Void Integral-field Treasury surveY,][]{Perez2024} is the first IFS survey designed to explore galaxies in cosmic voids. CAVITY is a legacy project of the Calar Alto observatory, which uses the Postdam MultiAperture Spectrometer (PMAS) IFS \citep{Roth2005, Verheijen2004} at the 3.5 m  telescope of the Calar Alto Observatory, the same as for the CALIFA survey, that allow us to cover the full optical extent of these galaxies in its field of view. The scope of CAVITY has expanded over time, including CO \citep{Rodriguez2024}, \ion{H}{I}, and deep imaging observations, to enable a more comprehensive study of void galaxies and maximise its scientific impact.

CAVITY IFS data has proven to be a fruitful source of scientific results to date \citep{Conrado2024,Sanchez2024,Bidaran2025}. In \citet{Conrado2024}, they analysed the stellar continuum of a set of void galaxy data cubes with full spectral fitting to retrieve their stellar population properties. Following this approach, they found out that galaxies in voids tend to be younger, have lower stellar mass surface density, and slightly higher SFR and sSFR than galaxies in filaments and walls. These effects were stronger for galaxies with lower stellar mass, in the outer parts of spiral galaxies, indicating less evolved discs, and on intermediate morphological types, suggesting a slower transition from star-forming to quiescent states in voids. Although analysing the stellar continuum gives us large amounts of information regarding the evolution of galaxies, stars are not the only physical elements that contribute to optical galaxy spectra. 

In this work, we extend the sample of CAVITY data cubes analysed in \citet{Conrado2024}, and enhance our study by inspecting the spatially-resolved SFR retrieved from the nebular information, using the extinction-corrected H$\alpha$ luminosity as the SFR tracer \citep{Kennicutt1998}. We evaluate the results for the whole galaxy (through global quantities), and as a function of galactocentric radius (with the radial profiles). We compare our results of galaxies in voids with a sample of galaxies in filaments and walls obtained from the CALIFA survey, matched in morphology and total stellar mass. Following the outlined methodology, we aim to assess the impact of the void environment on the SFR and sSFR of galaxies. The emission line properties analysed in this work (global and in 2D maps) will be available for public access with the CAVITY second data release\footnote{\url{https://cavity.caha.es/}}. 

The layout of this paper is the following. We summarise the observations, data reduction, and sample selection in Sections \ref{Section: observations} and \ref{Section: sample}, which are explained in more detail in \citet{Perez2024} and \citet{Garcia-Benito2024}. Section \ref{Section: method} contains the methodology. We present the main results in Section \ref{Section: results}, and compare them with a control sample of galaxies in filaments and walls in Section \ref{Section: comparison}. Section \ref{Section: discussion} discusses the results, and the summary is presented with the conclusions in Section \ref{Section: conclusions}. The masses calculated for this work were normalised with a Salpeter initial mass function (IMF). Throughout this work we assume a flat $\Lambda$CDM (Lambda cold dark matter) cosmology with $\Omega_M=0.3$, $\Omega_\Lambda=0.7$ and $H_0=70$ km s$^{-1}$ Mpc$^{-1}$.

\section{Observations and data reduction} \label{Section: observations}

The observations were conducted using the 3.5m telescope at the Calar Alto Observatory with the Potsdam Multi-Aperture Spectrometer \citep[PMAS;][]{Roth2005} in PPaK mode \citep{Verheijen2004}. The PPaK fibre bundle, consisting of 382 fibres, covers a $74"\times64"$ field of view, with each fibre having a diameter of $2.7"$ on the sky. The setup is arranged in a hexagonal pattern with a 60\% filling factor, and also includes six additional bundles of six fibres each to sample the sky background at the edges of the field of view. A three-position dithering pattern increases the filling factor to 100\%. This configuration produces datacubes of $78\times73$ pixels, where each pixel covers an area of $1"\times1"$. The observations were carried out in the low-resolution V500 mode ($R\sim850$ at 5000 \AA), with a full width at half maximum (FWHM) of $\sim6$ \AA \ and exposure times of either 1.5 or 3 hours, depending on the galaxy brightness. The wavelength range in this mode spans $3745-7500$ \AA, sampled in steps of 2 \AA. Further details on sample selection and observational strategy are described in \citet{Perez2024}, while data calibration and reduction is addressed in \citet{Garcia-Benito2024}.

The CAVITY pipeline builds upon the fundamental reduction
methodologies and protocols of the CALIFA survey \citep{Sanchez2016}, incorporating a new framework designed to meet the specific needs of the CAVITY survey. The reduction process accounts for the propagation of Poisson and read-out noise, as well as handling cosmic rays, bad CCD columns, and vignetting. It involves merging four FITS files into a single frame, performing pixel cleaning, fibre tracing, and stray-light correction. After spectral extraction and calibration, the data are homogenised to a common resolution, corrected for transmission variations, flux-calibrated, and adjusted for atmospheric extinction. Sky subtraction and datacube construction are then carried out, followed by corrections for differential atmospheric refraction and galactic extinction. Lastly, an absolute flux calibration anchored to SDSS DR16 \citep{Ahumada2020} is applied.

\section{Sample selection} \label{Section: sample}

The CAVITY parent sample includes 4866 galaxies distributed across 15 voids, carefully chosen from the \citet{Pan2012} void galaxy catalogue, which consists of 79947 galaxies within 1055 voids \citep{Perez2024}. These 15 voids were selected to encompass a variety of sizes and dynamical states. The parent sample is composed of galaxies from the nearby Universe, with redshifts between 0.01 and 0.05. It covers a broad range of stellar masses, from $10^{8.5}$ to $10^{11}$ M$_\odot$, estimated using the mass-to-light ratio derived from colours \citep{McGaugh2014}. The selected galaxies meet specific criteria: they reside deep within their respective voids (at least 80\% of the void’s effective radius), each void contains a minimum of 20 galaxies, and their distribution in right ascension ensures year-round observability from the Calar Alto Observatory.

Out of the 4866 galaxies of the CAVITY parent sample, 1115 are suitable to be observed with the described instrumental setting. The selected galaxies do not have excessively bright stars within the PMAS field, are of suitable size ($d_{25}\gtrsim 20"$) and surface brightness ($\lesssim 25 \ \text{mag}/"^2$) for this instrument, and exhibit intermediate inclinations (20 to 70 degrees). Among those observed up until the end of March 2025, we chose the ones with data quality sufficient for spatially resolved analysis \citep{Garcia-Benito2024}. We chose a lower limit of 50 Voronoi zones (see Section \ref{subsection: preprocessing}). Our sample consists of 216 datacubes of void galaxies from the CAVITY project. This sample includes the 100 selected galaxies for the CAVITY data release 1 \citep{Garcia-Benito2024}.

In our analysis, we incorporated four galaxies from the Void Galaxy Survey (VGS) by \citet{Kreckel2011, Kreckel2012}. These void galaxies were observed with the same instrument and setup as CAVITY in 2019 and 2020 as part of a pilot study. They do not belong to any of the 15 CAVITY voids but rather to other voids from the parent catalogue in \citet{Pan2012}. However, we included them given that they fulfill the same selection criteria as the CAVITY galaxies. As a result, our final selection consists of 220 void galaxies.

\begin{figure}
\centering
\includegraphics[width=0.45\textwidth]{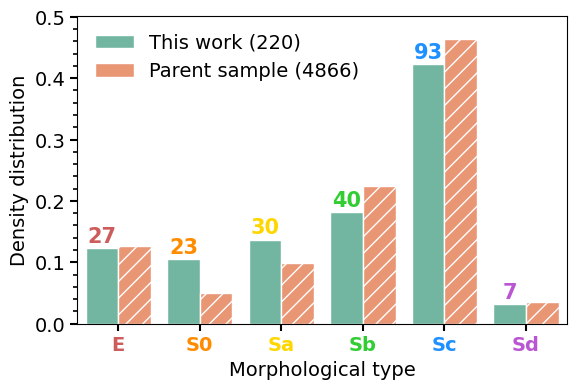}
\caption{Density distribution of morphological types in the CAVITY parent sample (4866 galaxies) and the selected ones (220 galaxies). The numbers above the bars indicate the number of galaxies in the selected sample of each morphological type.}
\label{fig:htype}
\end{figure}

For the morphological classification of CAVITY galaxies, we made a cross-match with the morphology catalogue from \citet{Dominguez-Sanchez2018}. To assign Hubble types, we used the T-Type parameter, which categorizes galaxies on a numerical scale.
Additionally, for early-type galaxies (T-Type < 0), we considered the probability of being S0 versus E, p(S0), provided in the same study. This additional parameter is used because the Hubble type classification model struggles to distinguish ellipticals from lenticulars. The p(S0) value is derived from a separate model specifically designed to address this limitation, independent of the model used for the T-Type classification. By incorporating this parameter, we ensure a more accurate morphological classification of our galaxies. The relationship between the parameters from \citet{Dominguez-Sanchez2018} and the Hubble types is detailed in Table \ref{table:TType-HType}. Some galaxies did not have a determined classification in the morphological catalogue, or we found them to be wrongly classified. In those cases (20), we determined visually their Hubble type based on their optical image. The morphological type distribution for both the parent sample and the selected subsample is presented in Figure \ref{fig:htype}.

\begin{table}[]
\centering
\caption{Hubble morphological classes as defined from the T-Type and probability of being S0, p(S0).}
\begin{tabular}{c|cccccc}
       & E              & S0           & Sa  & Sb  & Sc  & Sd  \\ \hline
T-Type & $<0$   & $<0$ & 0 -- 2 & 2 -- 4 &  4 -- 6 &  6 -- 8 \\
p(S0)  & $<0.5$ & $\geq0.5$    &     &     &     &    
\end{tabular}
\label{table:TType-HType}
\end{table}

\begin{figure}
\centering
\includegraphics[width=0.45\textwidth]{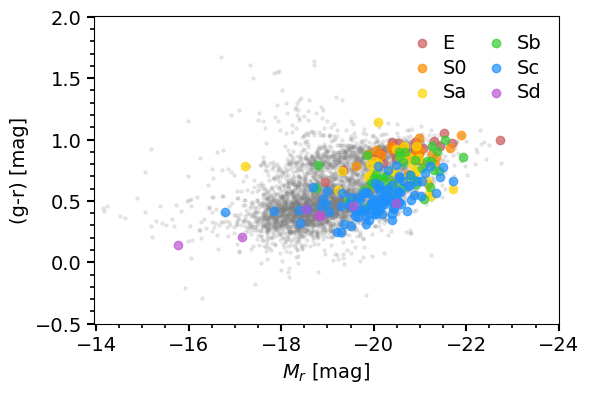}
\caption{Colour-magnitude diagram with the full CAVITY parent sample for galaxies with available Hubble types in \citet{Dominguez-Sanchez2018} (grey points) and the selected galaxies (coloured by morphological type).}
\label{fig:color-mag}
\end{figure}

Figure \ref{fig:color-mag} shows the colour-magnitude diagram of the parent sample and the selection for this work. It can be seen that our analysed sample covers the same area as the parent sample, with a smaller population at the fainter end. This is a result of the selection criteria: we select the brightest galaxies to ensure a sufficient signal-to-noise ratio (S/N) for reliable fitting at large galactocentric distances and to obtain spatially resolved information. If we compare with the diagram presented in Figure 2 of \citet{Conrado2024}, there is a noticeable improvement with respect to the coverage of the selected sample. This allows us to extract more robust conclusions.

\section{Analysis of the nebular emission lines} \label{Section: method}

To be able to obtain the SFR calculated from H$\alpha$ fluxes, there are several steps that must be followed. These are detailed in this section, where we explain the preprocessing necessary to obtain the emission line data cubes, their fitting, the reddening correction of the H$\alpha$ fluxes, and the calculation of the SFR.

\subsection{Preprocessing and stellar continuum subtraction} \label{subsection: preprocessing}

For a more accurate determination of the emission line fluxes, we must first subtract the contribution of the stellar continuum to the observed spectrum. To do so, we follow three steps: preprocessing the input data cube and segment it into Voronoi bins, perform full spectral fitting to obtain the synthetic continuum spectra, and dezonify the synthetic data cube to recover the pixel-by-pixel information.

We use \pycasso \ \citep{de_Amorim2017} as the pipeline to preprocess, fit, and analyse the 3D information of the stellar continuum in the CAVITY data cubes. Summarising, the preprocessing steps are the following: first, a spatial S/N mask is created to exclude all spaxels with S/N\textsubscript{threshold} < 3 per spaxel. The S/N is calculated as the median over the detrended standard deviation of the flux in a given rest-framed wavelength window. We choose this range to be $5650\pm60$ \AA, as this region of galaxy spectra is devoid of absorption or emission lines. If unwanted elements are present in the data cube, such as foreground stars or spurious sources, we apply a spatial mask to the corresponding regions to exclude them. The next step is binning the spaxels together, to make sure we have enough S/N to perform the stellar spectral fitting. The data cubes are segmented into Voronoi zones \citep{Cappellari2003}, with a S/N\textsubscript{target} = 20 per zone. Finally, the spectrum of each zone gets restframed and resampled to the range $3750-7000$ \AA, with $\Delta \lambda = 2$ \AA, and we apply a spectral mask to the emission lines and the flagged wavelengths during the reduction process. Once we have followed this steps, we fit the stellar contribution in each Voronoi bin with the full spectral fitting algorithm \starlight \ \citep{CidFernandes2005} in the wavelength range 3750 to 7000 \AA, as explained in \citet{Conrado2024}. During the fit, we use a combiation of the \citet{Vazdekis2010} (for ages $\leq32$ Myr) and \citet{Gonzalez-Delgado2005} (for ages < 63 Myr) SSPs, which assume a Salpeter IMF.

After the preprocessing and the fitting stages, we end up with a data products cube, which includes the resulting synthetic spectra for each of the zones. If we used such segmentation to do the emission line fitting, we would be losing information, given that the S/N of the emission lines is larger than the S/N of the stellar continuum. For this reason, we add an extra step before the continuum subtraction: we "dezonify" the synthetic data cube \citep{CidFernandes2013}. In each of the Voronoi zones, we calculate the contribution in light of every spaxel to the total of the zone, using the luminosity in the wavelength range $5635\pm45$ \AA \ as the reference. Then, we apply these weights to the synthetic spectrum of the zone, so that the flux level of each spaxel matches that of the corresponding observed spectrum. We thus obtain an emission-line data cube in spaxel resolution, free of segmentation and ready for fitting.

\subsection{Emission line measurement} \label{subsection: method EL fit}

In order to measure the emission lines, we use the SHerpa IFU line-fitting software \shifu\ (García-Benito et al., in preparation). \shifu\ is a Python-based package built on the sherpa modeling and fitting framework \citep{Freeman2001} within the CIAO environment \citep{Doe2007,Burke2023}. It provides a flexible interface for emission-line analysis, enabling standard Gaussian line fitting with a polynomial continuum, while allowing more complex configurations when required (e.g. blended lines or linked parameters such as velocity or FWHM). Uncertainties can be estimated using covariance, confidence interval, or Monte Carlo methods.

In our case, we fit the continuum-subtracted data cubes, grouping together the H$\beta$ and [\oiii]$_{4363}$ lines, as well as [\nii]$_{6548}$, H$\alpha$, and [\nii]$_{6584}$. Additionally, we tie the line widths and central velocities of H$\alpha$ to those of H$\beta$, and similarly couple the [\nii] doublet components to each other. Although the fit is carried out on the residual spectra, the continuum component is allowed to accommodate small deviations remaining from the stellar continuum subtraction. The continuum is modeled with a first-order polynomial, while emission lines are represented by single Gaussian profiles. When computing equivalent widths, the continuum level is evaluated from the original spectra.

To remove spaxels in which the measurements may be unreliable, we mask from the flux maps the spaxels with S/N$_{\text{H}\alpha}<1$, and with FWHM > 15 \AA. This number was chosen big enough to remove just the unphysical results.

\subsection{Reddening correction} \label{subsection: red_corr}

To account for extinction, we assume a foreground dust screen extinction model, with an extinction $A_\lambda$, that can be parametrised through the ratio between the H$\alpha$ and H$\beta$ emission line fluxes, also called the Balmer decrement. This quantity has a theoretical intrinsic value of $(F_{\text{H}\alpha}/F_{\text{H}\beta})_{0}=2.86$ when the physical conditions of the dust screen in the case of type B recombination correspond to an electronic temperature of 10$^4$K and an electron number density of $10^2 \ \text{cm}^{-3}$ \citep{Osterbrock1989}. We adopt the shape of the extinction law from \citet{Cardelli1989} with $R_V=3.1$ to calculate the extinction in the V-band ($A_V$), as well as $A_{\text{H}\alpha}$ to perform the reddening correction.

In some cases, issues when measuring $F_{\text{H}\beta}$ may result in an observed ratio $(F_{\text{H}\alpha}/F_{\text{H}\beta})_{obs}<2.86$, which is physically implausible with the stated assumptions. In such instances, the corresponding spaxels are masked in the extinction map, as their intrinsic flux cannot be reliably determined, along with spaxels where S/N$_{\text{H}\beta}<3$. Pixels that are unmasked in the $F_{\text{H}\alpha}$ map but masked in the $A_V$ map are left uncorrected to recover as much of the flux as possible. This can be interpreted as a lower limit for the magnitudes dependent on $F_{\text{H}\alpha}$.

\subsection{Star formation rate}

The amount of emission in H$\alpha$ correlates with the rate at which stars are being formed recently. From the dereddened luminosity in H$\alpha$, we can calculate the SFR corresponding to the last $\sim$10~Myr using the empirical relation derived in \citet{Kennicutt1998}:

\begin{equation}
    \text{SFR}[M_\odot/\text{yr}] = 7.9 \cdot 10^{-42}\cdot L_{\text{H}\alpha,\text{dered}}[\text{erg/s}].
\end{equation}

We consider for the estimation of the SFR those spaxels whose ionisation mechanism is characterised as star-forming following the criteria of the WHaN diagram \citep[][see Section \ref{Subsection: diagnostic diagrams}]{CidFernandes2010}. This means that we only keep the spaxels that satisfy that the equivalent width of H$\alpha$, EW(H$\alpha$), is larger than 3 \AA, and where the ratio in logarithm between the flux of the [\nii]$_{6584}$ and H$\alpha$ lines is lower than $-0.1$. The spaxels that do not fulfil these conditions will be masked for the SFR calculation. Because we focus in the radial structure of the SFR, we calculate the intensity of the SFR (or SFR surface density), which is the SFR per unit area, $\Sigma_{\text{SFR}}$, dividing the SFR by the area of the pixel.

The specific SFR, sSFR, is the SFR per unit stellar mass, which is the total stellar mass in the area of the galaxy that encompasses the spectrum (usually a spaxel), and can be easily obtained dividing the former by the stellar mass surface density:

\begin{equation}
    \text{sSFR} = \frac{\Sigma_{\text{SFR}}}{\Sigma_\star}\quad.
\end{equation}

We use the maps and global values of $\Sigma_\star$ calculated with the methodology applied in \citet{Conrado2024}, using the non parametric full spectral fitting algorithm \starlight\ \citep{CidFernandes2005} on the same IFU datacubes as in this work. These maps are segmented into Voronoi zones (see Section \ref{subsection: preprocessing}). For a more precise calculation of the sSFR, we "dezonify" the map of the stellar mass density (see Section \ref{subsection: method EL fit}).

\subsection{Global properties and radial profiles}

\begin{figure*}
\centering
\includegraphics[width=\textwidth]{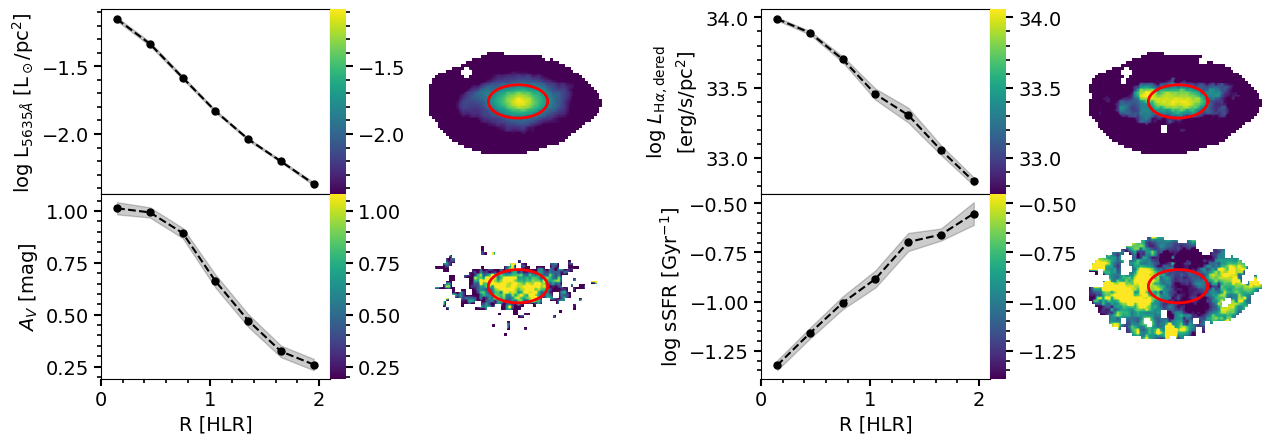}
\caption{Radial profiles and maps of some physical properties of CAVITY54706, one of the Sc galaxies in our sample. The shaded area around them mark the uncertainty, calculated as the standard deviation divided by the square root of the number of pixels taken into account. The red ellipse indicates its half-light radius. The colour bars share scale with their corresponding profile's y-axis. In the left, from top to bottom: Luminosity at the normalisation wavelength ($\lambda=5635$\AA) and nebular extinction. In the right column, from top to bottom: extinction-corrected H$\alpha$ luminosity and specific SFR.}
\label{fig:maps_profiles}
\end{figure*}

Once we have the maps of our desired properties, we calculate the global magnitudes and the radial profiles to be able to easily compare between galaxies. For those properties of extensive nature, such as the masses, fluxes, luminosities, or SFR, we can sum the value of all the non-masked pixels, to obtain a global value representative of the whole galaxy. In the case of extinction, we first correct each spaxel individually using its own $A_V$. The global $A_V$ is then calculated as the equivalent extinction that would convert the sum of all uncorrected spaxels into the sum of all extinction-corrected spaxels. This measurement can be used as a representative value of the galaxy's overall $A_V$.

To compare the spatially-resolved properties, we calculate the radial profiles, assuming axisymmetry and taking into account the position angle and ellipticity of each galaxy (derived from the flux map in the photometric r-band from the original datacube). To do so, we take the mean value inside rings of growing galactocentric radii, and normalise them by the HLR (the radius at which half of the light is contained, which is calculated at the same time as the position angle and ellipticity) to have all the profiles on a comparable physical scale. The error of the profile of a given property is obtained by taking the standard deviation inside each ring and dividing it by the square root of the number of non masked spaxels in it. To ensure that the measures are reliable, we mask the radii at which there are only 5 or less unmasked pixels inside the respective ring.
Figure \ref{fig:maps_profiles} displays the maps and corresponding radial profiles for several properties analyzed in this work, shown for an example void galaxy. A similar figure was presented in Figure 3 of \citet{Conrado2024} for the same CAVITY galaxy, focusing on its stellar population properties. Here, we update that content with the results from our analysis of the nebular component. Notably, the spatial resolution of the luminosity map at the normalization wavelength and of the sSFR map is improved as a result of the dezonification process.

\section{Emission line properties of CAVITY galaxies} \label{Section: results}

In this section, we analyse the resulting measurements of the emission line properties in our sample. First, making use of the diagnostic diagrams, we justify our criteria for masking spaxels or galaxies in which the measurements do not come from star formation processes. Then, we examine the behaviour of galaxies as whole through their global properties, and lastly study their distinct regions through their radial profiles.

\subsection{Diagnostic diagrams} \label{Subsection: diagnostic diagrams}

\begin{figure*}
\centering
\includegraphics[width=\textwidth]{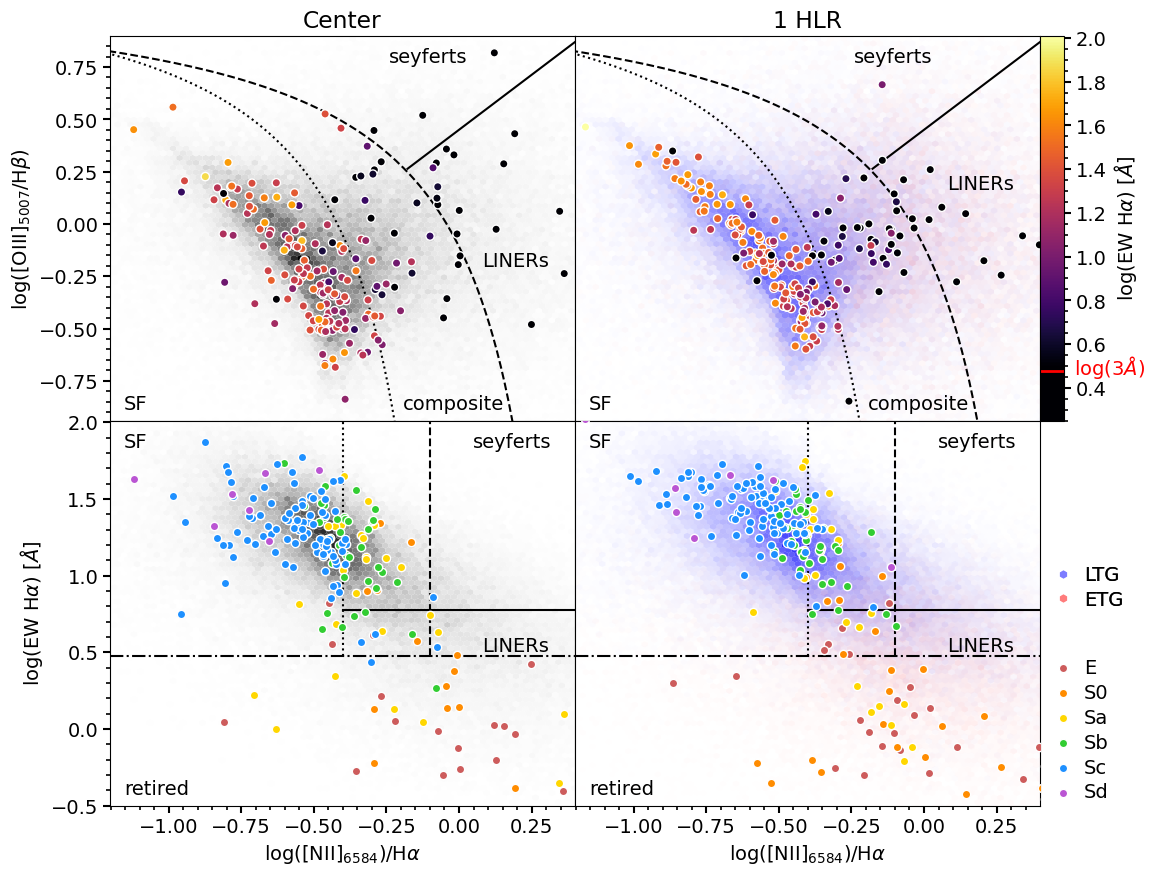}
\caption{BPT \citep[top, ][]{Baldwin1981} and WHaN \citep[bottom, ][]{CidFernandes2010, CidFernandes2011} diagnostic diagrams. The dots represent the flux values at the centre (left) or median values at a galactocentric distance of R = 1 HLR (right), coloured by the measurements of EW(H$\alpha$) (top) or by morphological type (bottom). The grey shadow in the left shows the distribution of the values of all pixels of all galaxies. On the right, the blue distribution corresponds to the late type galaxies (LTG; Sa, Sb, Sc or Sd), and the red to the early type galaxies (ETG; E or S0). The transparency of the colour of both morphological distributions is normalised to the total distribution, as in the panels of the left column. The dotted and dashed lines demarcate the limit between star formation and AGN following the criteria of \citet{Kauffmann2003} and \citet{Kewley2006}, respectively. The continuous line differentiates between Seyferts and LINERs \citep{Schawinski2007}. The dashed-dotted line in the two bottom panels mark delimit the retired galaxies (EW(H$\alpha$) < 3). Measurements below this limit are coloured black in the top panels.}
\label{fig:BPT}
\end{figure*}

The ionisation of the interstellar medium in not only caused by the radiation of OB-stars during star formation episodes, but from a variety of other physical mechanisms, e.g., shocks and hot low-mass evolved stars (HOLMES, \citealt{Flores-Fajardo2011,DAgostino2019}). A very common and straightforward way to classify the origin of the ionisation of the interstellar medium (ISM) is the use of diagnostics diagrams. Measuring the relative flux of some of the emission lines (usually comparing collisional lines with photoionised lines of the Balmer sequence, like H$\alpha$ or H$\beta$), we can know whether the source is star formation, active galactic nuclei (AGN), or a combination of both. 

One of such diagrams is the BPT diagram \citep{Baldwin1981}, which uses the ratio between [\oiii]$_{5007}$ and H$\beta$ against [\nii]$_{6584}$ and H$\alpha$ as diagnosis. The WHaN diagram is also popular \citep{CidFernandes2010, CidFernandes2011}, and owes its name to using the equivalent width of H$\alpha$ instead of the [\oiii] / H$\beta$ ratio. In both diagrams, we can classify our dominant ionising sources in star forming or non-stellar processes using the demarcation line of \citet{Kauffmann2003}. We can go further and divide the AGN population into Seyfert AGN or LINER (low ionization nuclear emission line regions) with the help of the \citet{Schawinski2007} criterion, and define an intermediate composite region between the \citet{Kauffmann2003} and \citet{Kewley2006} lines. Additionally, with the WHaN diagram, we can trace the distinction between spectra with ionised gas (from any of these mechanisms) and those that belong to regions or galaxies that are retired (that no longer have significant star formation) with the limit EW(H$\alpha$) < 3 \AA .

Figure \ref{fig:BPT} shows the diagnostic diagrams of the sample of CAVITY galaxies analysed in this work. 
From the BPT diagrams, we can see that there are almost no points in the Seyfert region, not even among the central spaxels. Most of them lay in the SF region, which we could expect given the predominance of late type galaxies (LTG; Sa, Sb, Sc or Sd) in our sample. From the colour of the points, which shows the EW(H$\alpha$), we can see that this value correlates with the ionisation mechanism: most of the points that lay in the SF region have EW(H$\alpha)$ larger or similar to 3 \AA. From the WHaN diagrams, we can see that all the galaxies that lie below the 3 \AA \ line are early type galaxies (ETG; E or S0), with the exception of some Sa spirals. The distribution of the spaxels that belong to LTG or ETG shows that the great majority of SF regions come from LTG. 

\subsection{Global properties}

\begin{figure}
\centering
\includegraphics[width=0.45\textwidth]{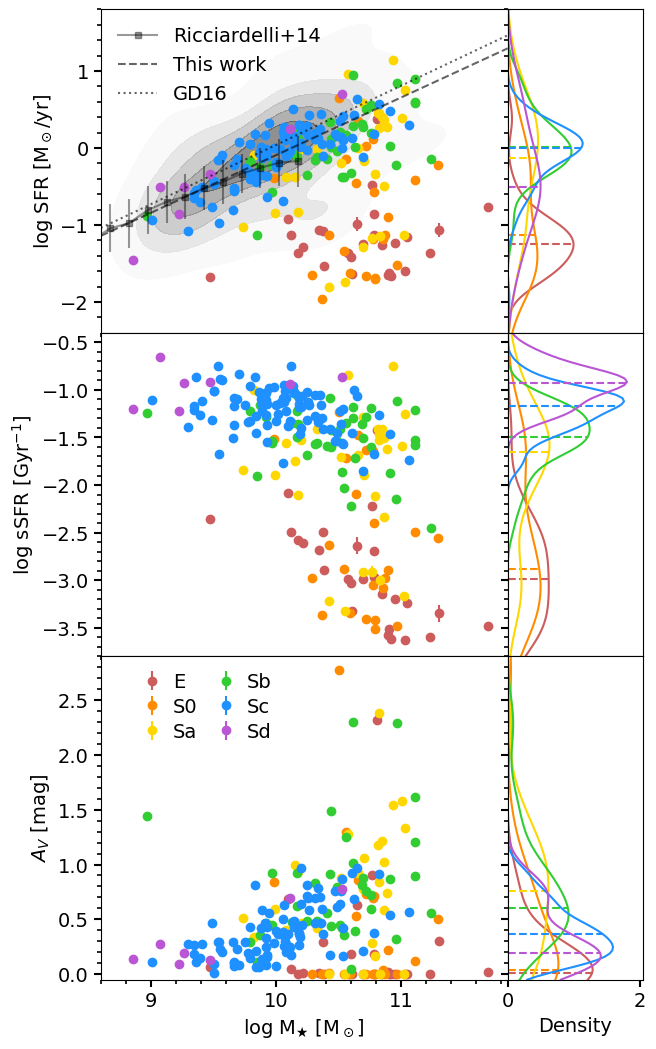}
\caption{Global properties coloured by morphological type. Each row shows a different property of each galaxy against their total stellar mass at the left, calculated in \citet{Conrado2024}, and the density distribution of each morphological type convolved with a gaussian kernel at the right, in the y-axis. The grey straight lines in the top left panel show the SFMS calculated by fitting the Sc and Sd galaxies of this work, derived from H$\alpha$ fluxes (dashed), and of the SFR from stellar population fitting of CALIFA galaxies in \citet{Gonzalez-Delgado2016} (dotted). The square dots reference the mean values obtained in \citet{Ricciardelli2014}, and the contours behind the dots the distribution from \citet{Rodriguez2024}, both for galaxies in voids. The horizontal dashed lines in the right panels mark the location of the median of each distribution.}
\label{fig:global_prop}
\end{figure}

The results that correspond to the global properties, calculated from the sum of the maps in all unmasked spaxels, are shown in Figure \ref{fig:global_prop}. The total stellar masses shown in the x-axis of panels in the left of the figure have been calculated from SED fitting using the same data cubes \citep{Conrado2024}. The panels at the right show the Gaussian smoothed distributions for each morphological type, with their median values marked with dashed lines.

The first row corresponds to the SFR. The relation between the SFR and the stellar mass $M_\star$ is called the star forming main sequence \citep[SFMS,][]{Brinchmann2004,Noeske2007}, and represents a universal scaling law for star-forming galaxies \citep{Speagle2014,Schreiber2015}. As their star formation is reduced during quenching, galaxies move from the SFMS (the blue cloud), passing through the green valley, to end up at the red sequence, which is populated by quiescent galaxies with very low SFRs \citep{Strateva2001}. This phenomenon can be seen in the top left panel, where the LTG are situated in a positive sequence, while the ETG lie much lower, with the exception of a few mixed ETG and LTG. This separation can be clearly seen in the distributions shown the top right panel, where the median values divide late- and early-type systems. We find the intersection between the distributions of spirals and early type galaxies (excepting the Sds) to be around $\text{log SFR}[M_\odot/\text{yr}]\sim-0.65$.

The grey lines above the scatter plot indicate the SFMS. The dashed line was obtained by fitting the Sc and Sd galaxies of this work, calculating the SFR from the H$\alpha$ luminosities, and using a Theil--Sen regressor. This technique is known for its robustness: it is a non parametric method which does not assume normal distributions, and has low sensitivity to outliers, given that it returns the median of the slopes determined by all pairs of sample points. The dotted line corresponds to the SFMS defined in \citet{Gonzalez-Delgado2016}, based on SFRs derived from the stellar continuum fitting of the Sc galaxies of the CALIFA survey. We use the latter as a way to illustrate a small systematic offset towards lower SFRs relative to the SED fitting-derived values. The median offset between the two lines in the mass range $10^9$~--~$10^{11.5} \ M_\odot$ (roughly the mass range of our sample) is 0.13 dex. To verify whether this offset is due to differences in methodology or sample selection, we recomputed the SFMS using the method of \citet{Conrado2024}, applied to the same galaxy sample and following an approach similar to \citet{Gonzalez-Delgado2016} (i.e., stellar-continuum-derived SFRs). This test confirms that the observed offset arises from the different SFR derivation methods, rather than from the galaxy selection. Moreover, we find the same behaviour when calculating the nebular sSFR in a comparable sample of galaxies located in filaments and walls (see Section~\ref{Section: comparison}). This topic will be further discussed in Section~\ref{subsec: SSP vs Ha}. To compare our results to the literature, we overplot the mean values of SFR for galaxies in voids from \citet{Ricciardelli2014}, as well as draw the contours of the distribution of the galaxies analysed in \citet{Rodriguez2024}. Both of these works use SFRs calculated from H$\alpha$ fluxes, enabling a direct comparison. Our results are in good agreement with these previous studies.

In the second row one can observe the results of the sSFR. The sSFR is a measurement of the relative rate at which galaxies are forming stars in the present, with respect to the past average rate. Similarly to the SFR, we can see a segregation with morphological types, where LTG reach higher values of sSFR in a sequence, while ETG populate the lower part of the panel, with higher dispersion. There is some mixing as well, with a few ETG higher up than their peers. For all morphological types, the median of the sSFR distribution of the rest of types are ordered accordingly with morphology. As it happened with the SFR, we find that the sSFR calculated from H$\alpha$ results in lower values than if it is calculated via stellar population analysis, even for the same data and sample (see Section \ref{subsec: SSP vs Ha}).

The last row shows the distribution of the global nebular extinction. We can see that, as expected, the extinction of ETG is very low, with a median value of 0 \citep{Gonzalez-Delgado2015}. In the case of the spirals, we see that the median values per morphological type are ordered, with earlier spirals having higher $A_V$. Similar results were found in \citet{Stasiska2001}, with larger extinctions for early spirals and a decrease towards later types. Additionally, we see an increase of $A_V$ of the ionised gas with total stellar mass for $M_\star>10^{9.6} \ M_\odot$.

\subsection{Radial profiles} \label{subsect: results radial profiles}

\begin{figure}
\centering
\includegraphics[width=0.45\textwidth]{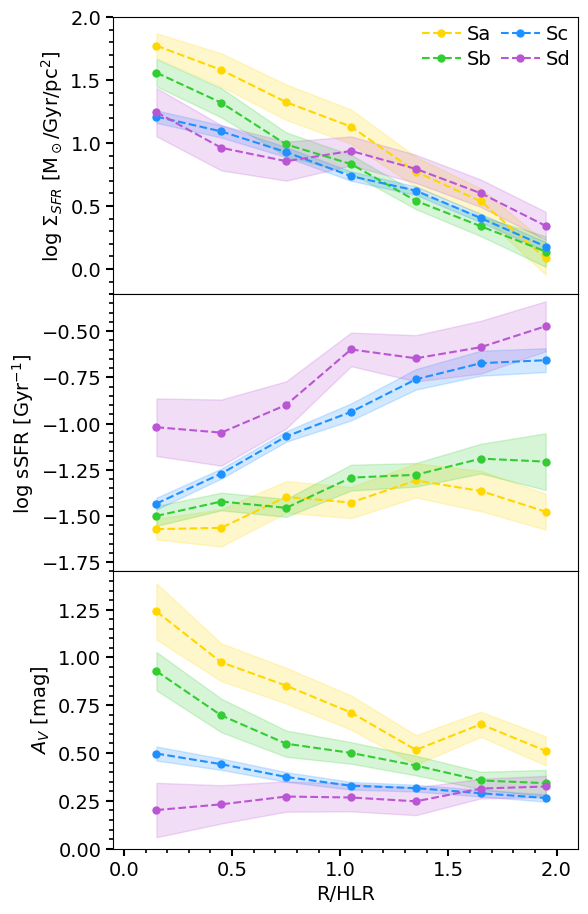}
\caption{Radial profiles of different emission line properties, stacked by morphological type. Each point represents the median value of the unmasked profiles at a given galactocentric distance, in bins of 0.3 HLR. Elliptical and lenticular galaxies are not represented given that most of the values of their profiles are masked. The shaded area marks the uncertainty, calculated as the standard deviation divided by the square root of the number of points in each R/HLR bin.}
\label{fig:radial_prof}
\end{figure}

Figure \ref{fig:radial_prof} shows the radial profiles stacked by morphological types, obtained by taking the median value of all the galaxies that belong to the same type at each galactocentric distance (normalised by the HLR of each galaxy). The coloured area around each line marks the error of the mean, calculated as the standard deviation divided by the square root of the amount of points at each radius from the centre. 
The stacked profile of each morphological type is calculated as the median of the unmasked values at each galactocentric radius, the same method used when calculating the profiles of the stellar population properties in \citet{Conrado2024}. Alternatively, we calculated them as the sum of the unmasked values divided by the total number of galaxies in each morphological bin, to also take into account the galaxies that have masked radial profiles. We found that, for the spiral galaxies, both methods yield very similar results. Therefore, for consistency, we adopt the same method as in our previous work.

In the case of the ETG, ellipticals and lenticulars, most of the pixels of their H$\alpha$ and H$\beta$ emission line maps are masked. We expect most of the ETG not to be actively forming stars, so this behaviour is consistent, given that we chose our masking criteria to only take star-forming regions into account. Consequently, the radial profiles are masked, so we are unable to calculate the median radial profile for their morphological types. For this reason, they are not represented in Figure \ref{fig:radial_prof}.

The top panel shows the profiles of the intensity of the SFR, the SFR per unit area.
We find that the profiles of all spiral types follow a common negative gradient, with similar values on their outer parts.
A similar behaviour is found for the $\Sigma_{\text{SFR}}$ derived from the stellar population analysis \citep{Conrado2024,Sanchez2024}, and is a consequence of spiral galaxies following the SFMS. The correlation of $M_\star$ with both the SFR and the size of the galaxy results in this invariance of the intensity of the SFR with the stellar mass \citep{Gonzalez-Delgado2016}. We can notice that the profile corresponding to the Sas deviate from the others in the inner regions, showing larger values of $\Sigma_{\text{SFR}}$. This could be due to a larger fraction of galaxies of said type to be masked, leading the median profile to be larger. We find that around 47\% of the $\Sigma_{\text{SFR}}$ profiles of Sa galaxies are masked in the innermost region, which is the largest fraction for all spiral types at any galactocentric radius. The reason for this higher masking fraction in their central regions varies from galaxy to galaxy: eight have passive stellar populations, one is classified as hosting an AGN, and five have fewer than five spaxels in the innermost radial bin, either due to their small size or high ellipticity. Something similar happens for he Sbs, which have a 12.5\% of galaxies masked in the innermost radial bin, being the second morphological type with a larger fraction of masked galaxies.

The profiles of the sSFR are in the middle panel. The profiles of Scs and Sds exhibit positive gradients with galactocentric radius, a sign of star forming discs, while the Sbs and Sas remain flat. Unlike the $\Sigma_{\text{SFR}}$, they scale with morphological type, with later types (Sc, Sd) reaching higher sSFR than earlier spiral types (Sa, Sb). We find that Sas and Sbs show similar profiles in the inner regions, with a large gap between their values and those of Scs. This difference grows with galactocentric radius, being the median difference between Sc/Sd and Sa/Sb types 0.13 dex at R/HLR = 0.15, and 0.66 dex at R/HLR = 1.95. As observed for the global properties, we find lower values for the $\Sigma_{\text{SFR}}$ and sSFR when calculating them from the H$\alpha$ luminosities and compare them to those obtained from the stellar population analysis.

The last panel represents the profiles of the extinction $A_V$.  Sds show a slightly positive gradient, while the rest of the spiral types exhibit negative gradients, that become steeper the earlier the morphological type.
Values are larger as well for earlier spirals, correlating with what was found for the global properties. Similar behaviour was found in \citet{Mingozzi2020}, which studied the profiles of nebular extinction not as a function of morphological type but stellar mass. They found more massive galaxies to have higher values and stronger negative gradients in their $A_V$ profiles. This result coincides with our findings, given the tendency of earlier types to be more massive (see Figure \ref{fig:Mdist_comparison}).

\section{Comparison with filaments and walls} \label{Section: comparison}

To assess the effect of the void environment on the SFR of galaxies, we compare our results with a similar sample of galaxies in other large-scale environments. In this section, we build a comparison sample of galaxies in filaments and walls from the CALIFA survey, and check whether their properties differ from the CAVITY galaxies we have analysed.

\subsection{Control sample selection} \label{subsect: control sample}

Following the same procedure as in \citet{Conrado2024}, we use galaxies in filaments and walls as a control sample (from now on we will refer to them as only filaments, for conciseness) from the CALIFA survey \citep{Sanchez2012}, observed with the same setup as the CAVITY data cubes, and analysed through the same methodology. We start with 653 Extended Data Release V500 datacubes \citep{Sanchez2023}, having removed those with unusual morphologies, such as ultra-compact systems, mergers, or irregular galaxies.

To build a control sample representative of intermediate-density environments, we make a cross match with two galaxy catalogues and exclude the galaxies that belong to voids (those with an effective radius from the void $\leq1$, from \citealt{Pan2012}) and to clusters (those that belong to groups with more than 30 members, from \citealt{Tempel2017}), the selection criteria used in \citet{Dominguez-gomez2022,Dominguez-Gomez2023a,Dominguez-Gomez2023b}. We did not keep galaxies in clusters as a third environmental group, as only 34 of them fulfilled the condition, making it impossible to build a sample comparable to galaxies in voids. After this second removal, we are left with 507 CALIFA galaxies located in filaments.

This way of selecting the filamentary environment is only a first approximation: filaments are complex structures that cover a vast range in density (5 orders of magnitude), ranging from the dense and heavy filamentary inflow arms to clusters, with a linear mass density of $\zeta_{\text{fil}}\sim10^{15} \ \text{M}_\odot$/Mpc, to underdense tendrils in the interior of voids, where $\zeta_{\text{fil}}$ can reach values as low as $10^{10} \ \text{M}_\odot$/Mpc \citep{Cautun2014}. The density distribution within them is inhomogeneous, but we find this approximation reasonable enough for the purposes of this work.

Many evolutionary properties of galaxies have a tight correlation with stellar mass \citep{Kauffmann2003, Baldry2006} and morphological type \citep{Perez2013, Gonzalez-Delgado2017, Lopez-Fernandez2018}. Therefore, to isolate the effect that the environment has on them, we must ensure that the two samples are well matched in mass and morphology. To do so, we select a pair for each galaxy in the void sample with the same morphological type and the closest stellar mass in the filament sample. In contrast to \citet{Conrado2024}, due to a larger void galaxy sample, we could not ensure no repetition of control sample galaxies if we wanted both distributions to be similar. This approach results in 151 unique CALIFA galaxies, each repeated at most two times within a given morphological bin, with most galaxies not repeated at all. The median and the maximum difference of log $M_\star/\text{M}_\odot$ between pairs is 0.01 and 0.34 dex, respectively. During the selection of the sample, we join the Sc and Sd galaxies ino a single morphological bin, due to the nonexistence of Sd galaxies in the control sample with stellar masses as low as for the void galaxies.

\begin{figure}
\centering
\includegraphics[width=0.45\textwidth]{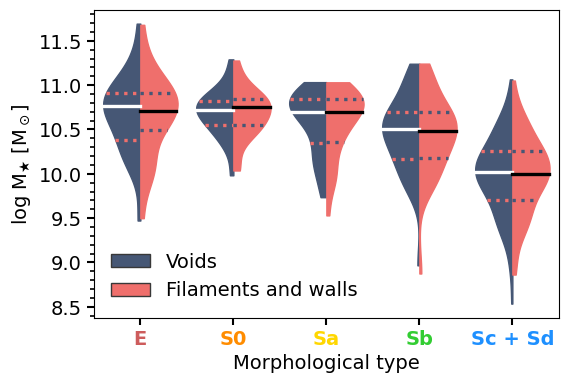}
\caption{Distribution of the total stellar mass by morphological types, for CAVITY void galaxies (in dark blue) and for CALIFA galaxies in filament or walls, matched in morphological type and total stellar mass (in red). The black and white lines mark the location of the median of each distribution, and the dotted lines their first and third quartiles. The edges of the distributions are cut to show the minimum and maximum values.}
\label{fig:Mdist_comparison}
\end{figure}

The total stellar mass distribution by morphological types of the final control sample and the CAVITY samples are shown in Fig. \ref{fig:Mdist_comparison}. We can see that, for every morphological type, both distributions are very similar, with median values almost identical (up to the second decimal). To have a quantitative measurement of the similarity of the distributions, we employ a two-sample Kolmogorov-Smirnov (K-S) test, a statistical test designed to determine whether two samples are drawn from the same distribution. This is done through the p-value, and the threshold is usually chosen as p $\geq0.05$ (which corresponds to a confidence level of 95\%). Therefore, we will consider the distributions to be different if p < 0.05. We find that for the total stellar mass distributions, as well as differentiating by morphological types, p > 0.75. Therefore, we conclude that the samples are well matched in terms of total stellar mass and morphology.

\subsection{Global properties}

\begin{figure}
\centering
\includegraphics[width=0.45\textwidth]{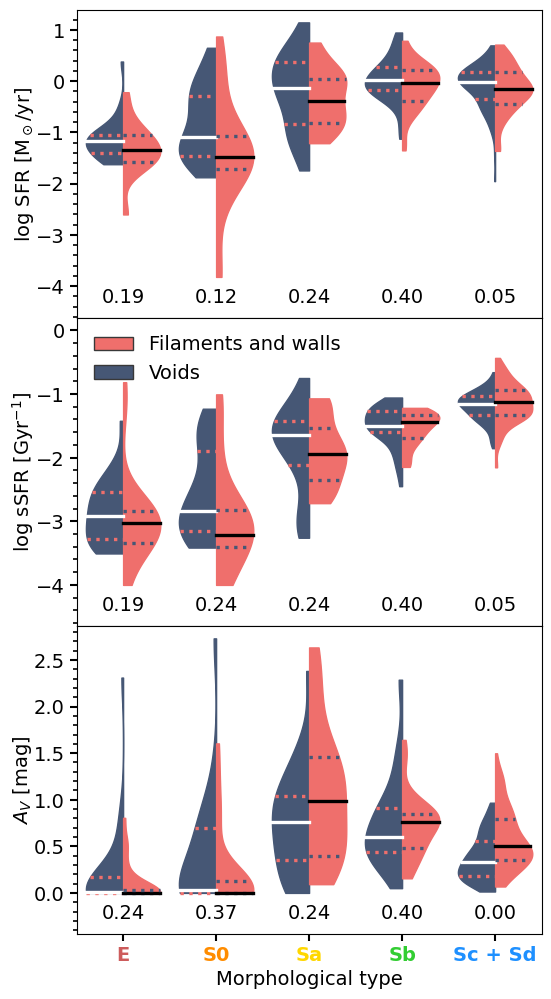}
\caption{Violin plots of the distributions of the global properties for void galaxies (in dark blue) and filament and wall galaxies (in red). From top to bottom, SFR, sSFR, and $A_V$. See Fig. \ref{fig:Mdist_comparison} for details. The numbers below every pair of distributions correspond to the p-value of the K-S test performed on them.}
\label{fig:violin_compar}
\end{figure}

Once we have built our control sample, we can compare their global properties, which can be seen in Figure \ref{fig:violin_compar} through their distribution segregated by morphological type. To gain insight into the statistical significance of these differences, we apply a K-S test to each pair of distributions, whose p-values are shown below each of them. To obtain a numerical value for comparison, we use two metrics. The first one is the D statistic from the K-S test, defined as the maximum vertical distance between the cumulative distributions. This number has a value between 0 and 1, and can give us an idea of the magnitude of the difference between the distributions. Table \ref{table:global_D} shows the D statistic for each of the compared properties, both for the whole sample and for categories of morphological type. The second metric is the difference of the median values ($\Delta=\Delta_{\text{void-fil}}$), which allow us to have a comparison in physical units, as well as to know which environmental sample shows higher values. The uncertainty of the median is calculated via bootstrapping, calculating the standard deviation of the bootstrap distribution, obtained by sampling with replacement from the original data 10000 times. The median values and their bootstrap uncertainty are shown in Table \ref{table:global_median}, for the different properties and morphological type bins. These two metrics are adequate to compare our results, given the non normality of the distributions of the analysed data, which can be appreciated in Figure \ref{fig:violin_compar}.

\begin{table}[]
\centering
\caption{D statistic of the K-test performed on the distributions of galaxies in voids and in filaments and walls, for each morphological type.} 
\begin{tabular}{c|ccc}
 Hubble Type & D(log SFR) & D(log sSFR) & D($A_V$) \\ \hline
E & $0.30 \ (0.19)$ & $0.30 \ (0.19)$ & $0.27 \ (0.24)$ \\ 
S0 & $0.39 \ (0.12)$ & $0.30 \ (0.24)$ & $0.25 \ (0.37)$ \\ 
Sa & $0.27 \ (0.24)$ & $0.27 \ (0.24)$ & $0.30 \ (0.24)$ \\ 
Sb & $0.20 \ (0.40)$ & $0.20 \ (0.40)$ & $0.20 \ (0.40)$ \\ 
Sc + Sd & $0.19 \ (0.05)$ & $0.19 \ (0.05)$ & $0.31$ (\textbf{0.00}) \\ 
Total & $0.11 \ (0.11)$ & $0.08 \ (0.39)$ & $0.14$ (\textbf{0.03}) \\ 
 \end{tabular}
\tablefoot{The values between brackets refer to the p-value. P-values lower than 0.05 are marked in bold.}
\label{table:global_D}
\end{table}

\begin{table}[]
\centering
\caption{Median difference of the global properties between galaxies in voids and in filaments and walls, for each morphological type.} 
\begin{tabular}{c|ccc}
 Hubble & $\Delta$ log SFR & $\Delta$ log sSFR & $\Delta$ $A_V$ \\ 
 Type & [M$_\odot$/yr] & [Gyr$^{-1}]$ & [mag]\\ \hline
E & $0.16 \pm 0.13$ & $0.10 \pm 0.18$ & $0.01 \pm 0.03$ \\ 
S0 & $0.45 \pm 0.36$ & $0.37 \pm 0.30$ & $0.03 \pm 0.19$ \\ 
Sa & $0.27 \pm 0.27$ & $0.30 \pm 0.16$ & $-0.24 \pm 0.24$ \\ 
Sb & $0.05 \pm 0.11$ & $-0.06 \pm 0.09$ & $-0.15 \pm 0.12$ \\ 
Sc + Sd & $0.15 \pm 0.05$ & $-0.04 \pm 0.05$ & $-0.16 \pm 0.06$ \\ 
Total & $0.12 \pm 0.07$ & $-0.02 \pm 0.06$ & $-0.10 \pm 0.06$ \\ 
 \end{tabular}
\tablefoot{Both the median values and the uncertainty are calculated via bootstrap (N resampling = 10000).}
\label{table:global_median}
\end{table}

The first panel of Figure \ref{fig:violin_compar} shows the distributions of the SFR. We find that, for all morphological types, galaxies in voids tend to have larger SFR than their counterpart in filaments.
The $\Delta$ is very close to 0 for the Sbs, which is the type that has the largest p-value, evidencing the similarity of the distributions. We find the largest differences for the S0s and Sas, which can also be appreciated though large D values, but p values larger than 0.05 for all the morphological types tell us about the similarity of the distributions.

The distributions of the sSFR are shown in the second panel. 
For the earlier morphological types (E, S0, Sa), the median sSFR is higher in voids than in filaments and walls. We find $\Delta<0$ for the latest types, although the values are close to 0 and the errors overcome them. The D values are very close to those obtained for the SFR, with earlier types having larger differences. As it happened with the SFR, we find all p values to be larger than 0.05, which tell us that the differences are not statistically significant.

For the bottom panel, which corresponds to the extinction, we find very similar values for the two environmental distributions for the ETG, and lower extinction in voids for the LTG. Although they show high D values, the $\Delta$ for ETG is very close to zero. For the LTG, we find $\Delta<0$, with Sas having the largest difference. Differences are also notable for the Scs and Sds, which have both large absolute $\Delta$ and D, and is the only case for which p < 0.05.

The p-values of the overall distribution, without separating by type, are 0.11, 0.39, and 0.03, for the log SFR, log sSFR, and $A_V$, respectively. Void galaxies tend to have larger SFR and sSFR, although the difference is not statistically significant. It is in the case of $A_V$, which is lower for void galaxies. This result is expected, given the dominance of Sc and Sd galaxies in our sample.

Consistently, the p-value for the Sc+Sd morphological bin is the smallest, even when other metrics show larger differences for other types. This may be due to sample size: the number of Sc+Sd galaxies is more than double that of the next most numerous bin (Sbs). Sample size can affect the precision of the p-value calculation, and we should take this into account when interpreting the results.

\subsection{Radial profiles}

\begin{SCfigure*}
\centering
\includegraphics[width=0.65\textwidth]{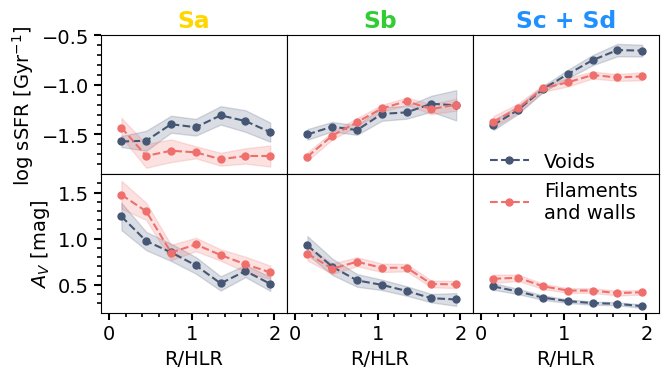}
\caption{Radial profiles of the sSFR (top row) and $A_V$ (bottom row) for spiral galaxies in voids and in filaments and walls, stacked by morphological type. Shaded areas mark the uncertainty (see Figure \ref{fig:radial_prof}).}
\label{fig:prof_compar}
\end{SCfigure*}

To check the effect of the void environment as a function of the region within the galaxy, we compare the radial profiles. They are shown in Figure \ref{fig:prof_compar}, for the sSFR and the $A_V$. We do not show the profiles of the SFR density, given the scale law explained in Section \ref{subsect: results radial profiles}. As we did when presenting the profiles of the void galaxies, we only show the results for the spiral morphological types.

The top three panels of the figure refer to the radial profiles of the sSFR, for our two environment samples and different morphological types. In the case of the Sas, we can see that void galaxies have larger sSFR at every galactocentric distance but the centre, with a difference of $\Delta=0.25$ dex at R = 1 HLR. This result was expected, given the large gap between the global sSFR medians for this morphological type. This difference is non existent for the Sbs, where the two profiles lay on top of each other, with $\Delta=-0.06$ dex at the same radius. In the case of Scs + Sds, the differences grow with galactocentric radius, with the outer parts being more affected by environment. For these types, the difference at 1 HLR is $\Delta=0.08$ dex, and $\Delta=0.26$ dex at 2 HLR. Differences between the trends observed in the radial profiles and the global properties can be explained by the fact that not all regions of a galaxy contribute equally when summed. Several factors can affect this weighting, including the larger number of pixels in the outer regions and the higher flux measured in the central areas.

At the bottom of the figure, we can see the radial profiles of the extinction. For the Sas in the first panel, we do not see a uniform trend with galactocentric radius. This is the morphological type for which we found the larger dispersion with the global properties. For the Sbs and Scs + Sds, the extinction of the void galaxies is lower at almost every galactocentric radii. The difference between the profiles at R = 1 HLR for the three morphological bins is $\Delta=-0.12$, $-0.14$, and $-0.13$ mag for the Sa, Sb, and Sc + Sd bins, respectively.

\section{Discussion} \label{Section: discussion}

In this section, we aim to achieve a deeper understanding of the presented results. First, we compare them with those obtained from analyses of the stellar continuum of the CAVITY data cubes \citep{Conrado2024}. Then, we focus on grasping the implications of the differences found on the early spirals, the population that has been found to be more affected by the environment in this study. For a comparative discussion in the context of galaxy evolution and comparison with previous works, check \citet{Conrado2024}.

\subsection{Comparison with stellar population properties} \label{subsec: SSP vs Ha}

Some galaxy properties can be obtained following different methods, as is the case of the SFR. There are several methods that can lead to its calculation, such as the measurement of ionised gas (performed in this work), or its derivation from the star formation history of the stellar population properties (as done in \citealt{Conrado2024}). 

From the results presented in Section \ref{Section: results}, it is evident that our \emph{nebular}, H$\alpha$-based SFR estimates are systematically lower than the \emph{stellar-population}-based values, both in the global and in the spatially resolved measurements. A test using the extended sample analysed with the same methodology as \citet{Conrado2024} (see Appendix~\ref{appendix: SPP}) yields consistent stellar-population-based SFRs and therefore similarly higher values than those obtained from the nebular tracers in this work.
In our case, the best-fit relation of the SFMS calculated from the H$\alpha$ luminosities (for Sc and Sd galaxies of the void + filament sample) is $\text{log SFR }[M_\odot/\text{yr}] = 0.75 - 7.58 \text{ log }M_\star [M_\odot]$. The same calculation for the same galaxy sample, but applied to the SFMS calculated from the stellar populations is $\text{log SFR }[M_\odot/\text{yr}] = 0.67 - 6.55 \text{ log }M_\star [M_\odot]$, which is above the previous relation for the totality of our $M_\star$ range.

Several factors may contribute to these differences, including the choice of stellar libraries and their comparison with the SFR$_{\text{H}\alpha}$ calibrator, effects on the latter that are not accounted for (such as metallicity), and the time-scale adopted to derive the SFR$_{\text{SSP}}$. A more detailed discussion of these issues can be found in \citet{Sanchez2019}.

Aside from the different values when using the two methods (analysing the H$\alpha$ luminosities in the current work and the stellar population properties in \citealt{Conrado2024}), we can see that the trends are consistent when comparing between large scale environments. 
In both studies, we find that void galaxies tend to have higher SFR and sSFR, especially for the Sas, although the difference is not statistically significant in the case of SFR$_{\text{H}\alpha}$ and sSFR$_{\text{H}\alpha}$. 
This tells us about the slower evolution from star-forming to passive in voids. Additionally, from looking at the profiles, we can see that the outer parts of LTG (Scs and Sds) are less evolved in voids, and therefore, more affected by the void environment. The concordance of these results allows us to state with more confidence the effect of the void environment on the evolution of galaxies.

\subsection{Slower evolution in void early spirals}

\begin{figure*}
\centering
\includegraphics[width=\textwidth]{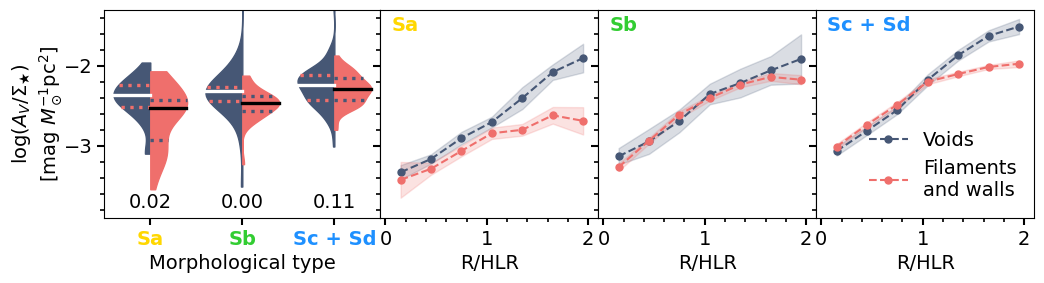}
\caption{Global values (left) and radial profiles (three right panels) of the $A_V/\Sigma_\star$ for the LTG in voids and in filaments and walls. Each dot represent the median value of the unmasked profiles at each galactocentric distance, and shaded areas around the radial profiles mark the uncertainty (see Figure \ref{fig:radial_prof}). The number below each violin plot in the left panel refers to the p-value of the K-S test done on the distributions.}
\label{fig:AvMdens_Sa}
\end{figure*}

The most pronounced environmental differences appear in Sa and S0 galaxies, with void galaxies exhibiting signs of a less evolved state, such as higher sSFRs. A key factor influencing a galaxy's evolutionary state is its gas content. Previous studies on void galaxies have yielded mixed results. \citet{Dominguez-gomez2022} and \citet{Rodriguez2024} report no significant difference in molecular gas fractions between galaxies in voids and in other environments. In the case of the atomic gas, in \citet{Pustilnik2016} they find larger fractions at fixed luminosity for galaxies in voids. In contrast, differences in the \ion{H}{I} fractions are not found in \citet{Rosas-Guevara2022}, using simulated data. We investigate this further using dust extinction as a proxy.

Given the established correlation between dust extinction and gas content \citep{Barrera-Ballesteros2020}, we use the extinction normalised by stellar mass surface density ($A_V/\Sigma_\star$) as a rough estimator of the gas fraction. These values are shown in Figure \ref{fig:AvMdens_Sa}, where we can see the distributions of the logarithm of the global $A_V/\Sigma_\star$ and its radial profiles for the LTG of our two environmental samples. We focus our analysis on Sa galaxies, as this morphological type exhibits the most pronounced differences. We can see that the gas fraction in void Sa galaxies is larger than for the control sample, with $\Delta=0.12\pm0.18$ dex (p < 0.05). These differences are larger in the outer parts of galaxies, with $\Delta=0.54$ and $\Delta=0.12$ dex at $\sim$1.5 and $\sim$0.5 HLR, respectively. This corresponds to an enhancement of 5\% in the global gas content for Sa void galaxies with respect to filaments walls, which decomposes into an 23\% increase in the outer parts (R $\sim$1.5 HLR) and a 4\% in the inner parts (R $\sim$0.5 HLR). In the later types one can see that the global differences (violin plots, left panel) are found in the same direction but are less pronounced (though still statistically relevant in the case of the Sbs), with $\Delta=0.14\pm0.07$ dex for the Sbs and $\Delta=0.05\pm0.03$ dex for the Scs and Sds. We find no discernible differences in the radial profiles of Sbs and the inner parts of Scs + Sds, but in the outer parts of the later galaxies (from R > 1 HLR) we find the same trend of void galaxies having more gas content, with $\Delta=0.39$ dex at $\sim$1.5 HLR (19\%).

The higher $A_V/\Sigma_\star$ in void Sa galaxies, particularly in their outer regions, implies a preservation of gas in low-density environments, which may sustain star formation longer and delay morphological transformation. Unlike the cited literature, which relies on integrated galaxy data, our trends require validation through spatially-resolved studies of neutral gas. This work is now in progress within the CAVITY collaboration, with related articles in preparation (Espada et al. in prep.).

These results have direct implications for the evolution of early-type spirals in voids, which predominantly populate the green valley. The differences found throughout this work can be read as dissimilarities in the quenching and morphological transformation time-scales in the two environments that we study. 

Several environmental mechanisms can influence galaxy evolution, driving morphological changes towards more bulge-dominated types (from spiral to elliptical) and leading to the cessation of star formation. These have a stronger effect in denser environments, which are as well more likely to host galaxy groups. While filaments represent a diverse range of densities, they are in average significantly denser than voids. Furthermore, filaments act as the primary pathways along which galaxies fall into clusters \citep{Kuchner2021}, experiencing important environmental pre-processing in the way \citep{Sarron2019}, caused by both their local and LSS environment.

In voids, early spirals likely experience slower quenching due to the absence of strong external mechanisms such as ram-pressure stripping or frequent minor mergers. In filaments, environmental pre-processing before infalling into clusters accelerates quenching, particularly in outer disks, leading to a faster transition toward passive states. Some evidence on the matter was found in \citet{Perez2025}, which found that ETG in voids are 10--20\% smaller than in denser environments, likely due to slower evolution and reduced number of minor mergers or accretion events.
In \citet{Oxland2024}, they analyse the effect that the pre-processing has on the morphological transformation and quenching time-scales. They find that, while both become shorter as the density of the environment increases, the later seems to be longer than the former, suggesting that quenching happens prior to morphological transformation. 
These results are consistent with our findings: star formation is suppressed more rapidly in filaments compared to voids, whereas morphological changes occur more slowly, reflecting their longer time-scale.

\section{Summary and conclusions} \label{Section: conclusions}

We analysed the ionised gas content of 220 IFU datacubes of galaxies in voids from the CAVITY survey, observed with the spectrograph PMAS/PPaK and the 3.5 m telescope in the Calar Alto Observatory. We measured their optical emission lines after continuum subtractions to obtain the flux maps. We then calculated maps of the SFR, sSFR, and $A_V$, and discussed them in terms of their global values (summing the maps) or their radial profiles (calculated by taking the median value in growing ellipses in accordance to each galaxy's ellipticity and position angle, and normalised by their HLR). The properties were then studied depending on their morphological type.

We repeated the same process for a control sample of CALIFA galaxies, selected for them to be in filaments or walls. We did so removing those that belong to a cluster \citep{Tempel2017} or void \citep{Pan2012}. To make the comparison fair, we build a control sample matched in morphology and stellar mass, by taking a pair filament galaxy for each void galaxy, allowing for repetition. In this process, we joined together the Scs and Sds, due to the lack of low mass filament galaxies to match the Sd void sample.

Our main results, based on the global and spatially resolved analysis, can be summarised as:

\begin{itemize}
  \item Differences in SFR and sSFR are not statistically significant. However, void galaxies consistently tend to have higher SFR, especially for the Sa and S0 morphological types.
  \item The extinction is found to be lower in void LTG. However, this difference is statistically significant only for the latest morphological types, as well as for the whole sample. It should be noted that this type is substantially more numerous than any of the others, and dominates the overall sample.
  \item The sSFR radial profiles show environmental differences in all spiral types but Sbs. The largest difference appears for Sas, with voids having larger sSFR at every galactocentric radius. The outer parts of Scs and Sds also show the same behaviour.
  \item The profiles of the $A_V$ do not show an uniform trend in the case of the Sas, but are lower for almost all galactocentric radii for the later morphological types.
\end{itemize}

This study shows new evidence of galaxies in voids to be affected by their environment, resulting in a slower evolution at fixed morphological type and stellar mass. This effect is especially strong for the early spirals, where we also find evidence of a higher gas fraction in voids, especially in their outer regions. This can serve as proof of galaxies in transition between star forming and passive to be more influenced by the large-scale environment. This influence is also present in the outer parts of late-type spirals, which leads to spiral disc in void galaxies to be less evolved.

These results are consistent with those obtained with the analysis of the stellar population properties \citep{Conrado2024}, allowing us to make stronger assertions about the effect of the void environment in the evolution of galaxies.

\section*{Data availability}

The tables with the global values for the fluxes of H$\alpha$ and H$\beta$, SFR and sSFR calculated from H$\alpha$ luminosities, as well as the global stellar populations properties for the updated sample calculated as in \citet{Conrado2024}, are only available in electronic form at the CDS via anonymous ftp to \url{cdsarc.u-strasbg.fr} (130.79.128.5) or via \url{http://cdsweb.u-strasbg.fr/cgi-bin/qcat?J/A+A/}. The 2D maps of these properties, and the maps of the dezonified stellar mass density, will be made publicly available in the CAVITY data base with its second DR, scheduled for the summer of 2026, at \url{https://cavity.caha.es/}.

\begin{acknowledgements} 
AC, RGB, and RGD acknowledge financial support from the Severo Ochoa grant CEX2021-001131-S funded by MCIN/AEI/ 10.13039/501100011033, and PID2022-141755NB-I00. 
This paper is based on observations collected at the Centro Astron\'omico Hispano en Andaluc\'ia (CAHA) at Calar Alto, operated jointly by Junta de Andaluc\'ia and Consejo Superior de Investigaciones Cient\'ificas (IAA-CSIC), under the CAVITY legacy project. The CAVITY project acknowledges financial support by the research projects AYA2017-84897-P, PID2020-113689GB-I00, PID2020-114414GB-I00, and PID2023-149578NB-I00 funded by the Spanish Ministry of Science and Innovation (MCIN/AEI/10.13039/501100011033) and by FEDER/UE; the project A-FQM-510-UGR20 funded by FEDER/Junta de Andaluc\'ia-Consejer\'ia de Transformaci\'on Econ\'omica, Industria, Conocimiento y Universidades/Proyecto; by the grants P20-00334 and FQM108, funded by Junta de Andaluc\'ia; and by Consejer\'ia de Universidad, Investigaci\'on e Innovaci\'on (Junta de Andalucía) and Gobierno de España and European Union NextGenerationEU through grant AST22\_4.4.
IMC acknowledges support from ANID programme FONDECYT Postdoctorado 3230653 and ANID, BASAL, FB210003.
TRL acknowledges support from Ram\'on y Cajal fellowship (RYC2023-043063-I, financed by MCIU/AEI/10.13039/501100011033 and by the FSE+).
GTR acknowledges financial support from the research project PRE2021-098736, funded by MCIN/AEI/10.13039/501100011033 and FSE+.
MAF and PVG acknowledge support from the Emergia program (EMERGIA20\_38888) from Consejer\'ia de Universidad, Investigaci\'on e Innovaci\'on de la Junta de Andaluc\'ia.
SBD acknowledges financial support from the project PID2021-122544NB-C43.
YGK acknowledges financial support from PREP2023-001684 funded by MCIU/AEI/10.13039/501100011033 and the FSE+.
AZLA gratefully acknowledges the support provided by the postdoctoral program (POSDOC) of UNAM (Universidad Nacional Autónoma de México).
PVG acknowledges that the project that gave rise to these results received the support of a fellowship from “la Caixa” Foundation (ID 100010434). The fellowship code is B005800.
This work has been supported by the Agencia Estatal de Investigación Española (AEI; grant PID2022-138855NB-C33), by the Ministerio de Ciencia e Innovación (MCIN) within the Plan de Recuperación, Transformación y Resiliencia del Gobierno de España through the project ASFAE/2022/001, with funding from European Union NextGenerationEU (PRTR-C17.I1), and by the Generalitat Valenciana (grant CIPROM/2022/49).
MIR acknowledges the support of the Spanish Ministry of Science, Innovation and Universities through the project PID-2021-122544NB-C43.
This research made use of {\sc astropy}, a community-developed core {\sc python} (\url{http://www.python.org}) package for Astronomy \citep{Astropy2013,Astropy2018,Astropy2022};
{\sc matplotlib} \citep{matplotlib2007}; {\sc numpy} \citep{numpy2020}; {\sc scipy} \citep{scipy2020}; {\sc pandas} \citep{pandas2020}; and {\sc seaborn} \citep{seaborn2021}.
This research has made use of the NASA/IPAC Extragalactic Database, operated by the Jet Propulsion Laboratory of the California Institute of Technology, un centract with the National Aeronautics and Space Administration. Funding for SDSS-III has been provided by the Alfred P. Sloan Foundation, the Participating Institutions, the National Science Foundation, and the U.S. Department of Energy Office of Science. The SDSS-III Web site is \url{http://www.sdss3.org/}. The SDSS-IV site is \url{http://www.sdss.org}.
\end{acknowledgements}

\bibliographystyle{aa}
\bibliography{bibliography}

\begin{appendix}

\section{Stellar population properties of the extended sample} \label{appendix: SPP}

We present the results obtained for the stellar population properties using the extended CAVITY sample of 220 galaxies, compared with the same control sample derived as in this work (see Section \ref{subsect: control sample}). The details of the methodology and the previous results with the sample of 118 galaxies are described in \citet{Conrado2024}.

\begin{figure*}[!b]
\centering
\includegraphics[width=\textwidth]{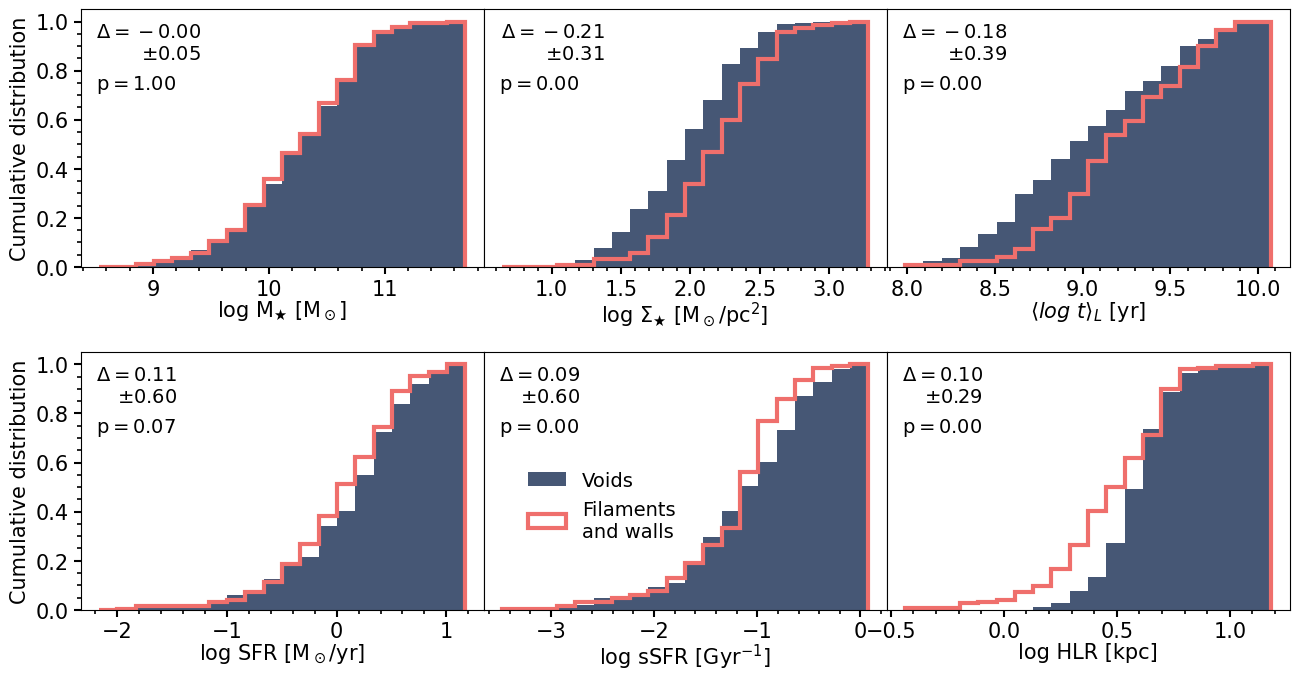}
\caption{
  Cumulative histograms of the global stellar population properties and HLR for the updated sample of voids (in dark blue) and filaments and walls (overlined in red). The properties plotted are, from left to right, stellar mass, stellar mass density, light-weighted age, SFR, sSFR and half-light radius. The median and the standard deviation of the subtraction between voids and filaments and walls ($\Delta=\Delta_{\text{void-fil}}$) is shown in the top left or right corner of each panel. Below, it is shown the p-value resultant of a two-sample K-S test applied to the distributions of the two samples.
}
\label{fig:globalprop_SP}
\end{figure*}

Figure \ref{fig:globalprop_SP} shows the cumulative histograms of the global stellar population properties with the extended sample. On the corner of each panel, we have the measurements of the median and standard deviation of the subtraction of the values of each pair ($\Delta=\Delta_{\text{void-fil}}$), as well as the p-value resulting from the K-S test. Although this way of measuring the uncertainty is not optimal for non normal distributions such as ours, we choose to display them this way for consistency with how they were calculated in \citet{Conrado2024}. We find that the differences between the distributions of the two environmental samples are statistically significant for all of our analised properties. The only exceptions are the stellar mass, which must be similar by construction, and the SFR, with a p-value of 0.07. These differences are similar than how they were with the more limited sample (which can be seen through the values of $\Delta$). We conclude that galaxies in voids are, in general, less dense, younger and have higher SFR and sSFR than galaxies in filaments and walls for the same stellar mass.

If we segregate by morphological type, we find that for most attributes and morphological bins the differences stand accordingly to the global $\Delta$. In the case of ellipticals we find $\Delta(\text{log sSFR})>0$, caused by only one pair of galaxies not having masked SFR. For the Sbs, we obtain $\Delta(\text{log SFR})>0$ and $\Delta(\text{log sSFR})>0$, which cannot be explained by number statistics alone. The largest differences are systematically found for the S0s, followed by the Sas. Checking the p-values obtained by testing the morphological segregated distributions we find that for the $\Sigma_\star$ every type but the ellipticals shows p-values < 0.05. In the case of the ages, this happens for every type but the Es and the S0s. In the case of the SFR no morphological bin has p < 0.05, threshold that is surpassed by the Scs + Sds for the sSFR and the HLR.

\begin{figure*}[!t]
\centering
\includegraphics[width=\textwidth]{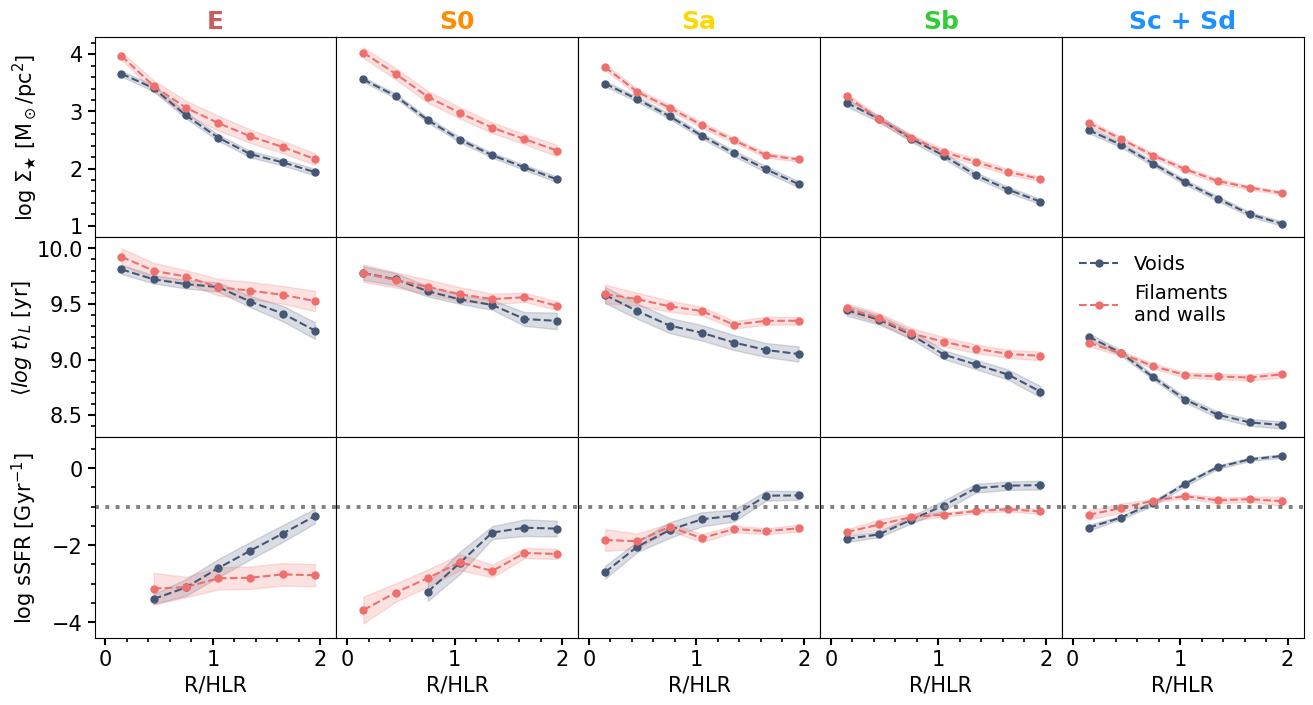}
\caption{Radial profiles of the stellar population properties for the updated sample of galaxies in voids and in filaments and walls, stacked by morphological types in different panels. From top to bottom, stellar mass density, age, and sSFR. The dotted grey line in the lower row of panels refers to the limit between star-forming and quiescent from \citet{Peng2010}.}
\label{fig:profiles_SP}
\end{figure*}

The comparison of the profiles of the stellar mass density, ages, and sSFR for the expanded sample are shown in Figure \ref{fig:profiles_SP}. Comparing to what was obtained for the smaller sample in \citet{Conrado2024}, we can see that the tendencies are reproduced. We find that the stellar mass density of void galaxies tend to be lower, trend that is followed more strongly in the outer parts of later type galaxies. For the ages, we see that the outer parts of galaxies are more affected by the environment for every morphological type. This gap towards younger ages in voids is specially evident for Sas, at almost all radii, and Scs + Sds. We find this differences to be larger than in \citet{Conrado2024}. The last row shows the profiles of the sSFR, which show a very similar trend to that of the ages. The outer parts of all morphological types result in larger sSFR in void galaxies. These results are also consistent to what we obtained for the smaller sample, with larger differences between environments. 


\balance

\FloatBarrier

\section{Environmental classification based on CosmicFlows} \label{appendix: vweb}

There are several ways in which the belonging of a galaxy to a LSS environment can be determined. In this paper, the LSS definition is based on galaxy redshift surveys, but there are other observables that can serve for the same purpose, such as galaxy peculiar velocities, which can classify the LSS according to the direction of the velocity field in 3D. Such is the case of CosmicFlows \citep{Tully2016}, from which we can classify volumes as voids (where the velocity field is expelling matter along the three directiont), sheets (which expand in two directions and contract in one), filaments (expanding in one direction and contracting in two), and knots (infalling in the three directions).

This survey has already been used in relation to the CAVITY project in \citet{Courtois2023}, where they analysed the dynamical state of seven of the CAVITY voids. In this case, we use it for the characterisation of the control sample, as an alternative methodology to appeal to catalogues based on galaxy redshift surveys. We classify all the 653 CALIFA galaxies depending on their LSS environment following this procedure, which results in 38 galaxies in voids, 42 in knots (clusters), 284 in sheets (walls), and 289 in filaments, with the two latter grouped together into the same category. Then, we build a control sample of galaxies in filaments and walls matched in morphological type and mass (see Section \ref{subsect: control sample}), which leads to a sample of 164 unique CALIFA galaxies, repeated at most two times. 

The comparison of the global properties of the two environmental samples is summarised in Tables \ref{table:global_D_web} and \ref{table:global_median_web}. The first table shows the D statistic and p-values of the K-S test performed on the two environmental sample. The second table shares the difference of the median values of the two distributions and its uncertainty, calculated using bootstrapping. We can see there is a strong correlation with the results of the original control sample (categories based on galaxy redshift catalogues). This time, more categories of morphological type have p < 0.05, and the same tendencies are found: galaxies in voids tend to have larger SFR, sSFR and lower $A_V$. These differences are more notable for the S0s and Sas in the case of the SFR (p < 0.05) and sSFR (not statistically significant), and for the Sas and Scs + Sds in the case of the $A_V$. We also find p < 0.05 by comparing the SFR and $A_V$ of the total sample.

\begin{table}[h!]
\centering
\caption{D statistic of the K-S test performed on the distributions of galaxies in voids and in filaments and walls (classified with CosmicFlows), for each morphological type.} 
\begin{tabular}{c|ccc}
 Hubble Type & D(log SFR) & D(log sSFR) & D($A_V$) \\ \hline
E & $0.26 \ (0.33)$ & $0.22 \ (0.53)$ & $0.15 \ (0.94)$ \\ 
S0 & $0.43$ (\textbf{0.02}) & $0.35 \ (0.12)$ & $0.26 \ (0.42)$ \\ 
Sa & $0.23 \ (0.39)$ & $0.23 \ (0.39)$ & $0.33 \ (0.07)$ \\ 
Sb & $0.25 \ (0.16)$ & $0.28 \ (0.10)$ & $0.28 \ (0.10)$ \\ 
Sc + Sd & $0.19 \ (0.05)$ & $0.22$ (\textbf{0.02}) & $0.30$ (\textbf{0.00}) \\ 
Total & $0.14$ (\textbf{0.03}) & $0.10 \ (0.27)$ & $0.17$ (\textbf{0.00}) \\ 
 \end{tabular}
\tablefoot{The values between brackets refer to the p-value. P-values lower than 0.05 are marked in bold.}
\label{table:global_D_web}
\end{table}

\begin{table}[h!]
\centering
\caption{Median difference of the global properties between galaxies in voids and in filaments and walls (classified with CosmicFlows), for each morphological type.} 
\begin{tabular}{c|ccc}
 Hubble & $\Delta$ log SFR & $\Delta$ log sSFR & $\Delta$ $A_V$ \\ 
 Type & [M$_\odot$/yr] & [Gyr$^{-1}]$ & [mag]\\ \hline
E & $0.20 \pm 0.15$ & $0.19 \pm 0.19$ & $0.01 \pm 0.04$ \\ 
S0 & $0.35 \pm 0.32$ & $0.37 \pm 0.28$ & $-0.01 \pm 0.19$ \\ 
Sa & $0.34 \pm 0.26$ & $0.36 \pm 0.21$ & $-0.35 \pm 0.21$ \\ 
Sb & $0.15 \pm 0.11$ & $0.05 \pm 0.09$ & $-0.14 \pm 0.13$ \\ 
Sc + Sd & $0.14 \pm 0.05$ & $-0.06 \pm 0.04$ & $-0.23 \pm 0.06$ \\ 
Total & $0.16 \pm 0.06$ & $0.02 \pm 0.07$ & $-0.17 \pm 0.05$ \\ 
 \end{tabular}
\tablefoot{Both the median values and the uncertainty are calculated via bootstrap (N resampling = 10000).}
\label{table:global_median_web}
\end{table}

\begin{SCfigure*}[]
\centering
\includegraphics[width=0.65\textwidth]{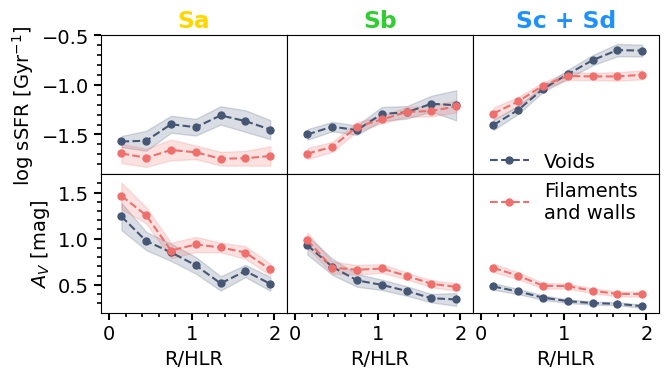}
\caption{Radial profiles of the sSFR (top row) and $A_V$ (bottom row) for spiral galaxies in voids and in filaments and walls, categorised using CosmicFlows, stacked by morphological type. Shaded areas mark the uncertainty (see Figure \ref{fig:radial_prof}).}
\label{fig:prof_compar:vweb}
\end{SCfigure*}

To show the comparison of the spatially-resolved properties with this control sample, we plot the radial profiles of the sSFR and $A_V$ in the same way we did with the previous results, which is shown in Figure \ref{fig:prof_compar:vweb}. We find almost identical profiles if we compare with the previous version. Again, we find the sSFR of Sa void galaxies to be larger in all galactocentric distances, which is found also in the outer parts of void Scs and Sds. For the extinction, it tend to be lower in voids at almost all distances for the different types of LTG.

The similarity of the results builds confidence not only in the physical conclusions of this work, but also on the methods that are commonly used to identify LSS environments, which have been proven to converge to the same diagnosis.


\end{appendix}
\end{document}